\documentclass[conference]{IEEEtran}
\IEEEoverridecommandlockouts
\usepackage{cite}
\usepackage{amsmath,amssymb,amsfonts}
\usepackage{algorithmic}
\usepackage{graphicx}
\usepackage{textcomp}
\usepackage{xcolor}
\usepackage{subfig}
\usepackage{graphicx}
\usepackage[export]{adjustbox}
\usepackage{longtable}
\usepackage{multicol}
\usepackage{multirow}
\usepackage{graphicx}
\usepackage{lscape}
\usepackage{url}

\def\BibTeX{{\rm B\kern-.05em{\sc i\kern-.025em b}\kern-.08em
    T\kern-.1667em\lower.7ex\hbox{E}\kern-.125emX}}

\usepackage{fancyhdr}
\begin{document}

\title {Hybridized Convolutional Neural Networks and Long Short-Term Memory for Improved Alzheimer's Disease Diagnosis from MRI Scans}

\author{
\IEEEauthorblockN{
Maleka Khatun\textsuperscript{1},
*Md Manowarul Islam\textsuperscript{1}, 
Habibur Rahman Rifat\textsuperscript{1}, 
Md. Shamim Bin Shahid\textsuperscript{1},\\
Md. Alamin Talukder\textsuperscript{2}, 
*Md Ashraf Uddin\textsuperscript{3}
}
\IEEEauthorblockA{\textsuperscript{1}Department of Computer Science and Engineering, Jagannath University, Dhaka, Bangladesh}
\IEEEauthorblockA{\textsuperscript{2}Department of Computer Science and Engineering, \\International University of Business Agriculture and Technology, Dhaka, Bangladesh}
\IEEEauthorblockA{\textsuperscript{3}School of Information Technology, Deakin University, Geelong Waurn Ponds Campus, Australia}
}

\maketitle
\begin{abstract}

Brain-related diseases are more sensitive than other diseases due to several factors, including the complexity of surgical procedures, high costs, and other challenges. Alzheimer's disease is a common brain disorder that causes memory loss and the shrinking of brain cells. Early detection is critical for providing proper treatment to patients. However, identifying Alzheimer's at an early stage using manual scanning of CT or MRI scans is challenging. Therefore, researchers have delved into the exploration of computer-aided systems, employing Machine Learning and Deep Learning methodologies, which entail the training of datasets to detect Alzheimer's disease. 
This study aims to present a hybrid model that combines a CNN model's feature extraction capabilities with an LSTM model's detection capabilities. This study has applied the transfer learning called VGG16 in the hybrid model to extract features from MRI images. The LSTM detects features between the convolution layer and the fully connected layer. The output layer of the fully connected layer uses the softmax function. 
The training of the hybrid model involved utilizing the ADNI dataset. The trial findings revealed that the model achieved a level of accuracy of 98.8\%, a sensitivity rate of 100\%, and a specificity rate of 76\%. The proposed hybrid model outperforms its contemporary CNN counterparts, showcasing a superior performance.

\end{abstract}

\begin{IEEEkeywords}
Alzheimer’s disease, CNN, Inception V3, Resnet50, Vgg16, Deep learning, LSTM, MRI image.
\end{IEEEkeywords}

\section{Introduction}
Alzheimer's disease, a chronic neurological ailment, is recognized for its progressive deterioration of memory and cognitive capacities \cite{rana2023robust}. This condition stands as the primary identified cause of dementia among the elderly population. Dementia, a costly affliction, incurred a global expense of \$604 billion in 2010 alone. Notably, 60\% to 80\% of individuals afflicted by dementia fall under the category of Alzheimer's disease.\cite{zhang2017advancing}. The scientific name of Alzheimer's disease is derived from the renowned physician, Dr. Alois Alzheimer.\cite{alzheimer}. In 1906, Dr. Alzheimer noticed some changes in the brain tissue of a dead woman who was suffering from an unusual mental illness. Dr. Alzheimer observed that there were several aberrant plaques and twisted bundles of fibers in her brain. This damage initially occurs in the entorhinal cortex and hippocampus and then affects the cerebral cortex \cite{bhattacharjya2024exploring}. Alzheimer's disease must be identified in its earliest stages in order to prevent irreversible brain damage in the patient. In the case of individuals aged 65 or older, it becomes hazardous and occasionally fatal.


Presently, the global impact of Alzheimer's disease spans over 55 million individuals, with predictions foreseeing a surge to 75 million by 2030 and a staggering 139 million by 2050\cite{stat}. Astonishingly, a new case joins these ranks every 3.2 seconds\cite{stat_dhs}. Despite comprehensive investigations, the precise etiology of Alzheimer's remains elusive. Rather than pinpointing a singular cause, scientists hypothesize that a blend of genetic, lifestyle, and environmental factors collaboratively influence the brain, gradually precipitating the onset of this condition.

The absence of a cure for Alzheimer's disease is a significant cause of concern. Several reasons make it difficult to use machine learning to create a model of Alzheimer's disease (AD) development. To begin, there is a multimodal quality to AD data\cite{zhang2017advancing}; that is, the data for any one patient might come from a number of different places. These include a wide variety of information sources, including but not limited to MRI scans, cerebrospinal fluid (CSF) measurements, neuropsychological tests, blood pressure readings, heart rate readings, cognitive scores, laboratory test results, neurological assessments, PET scans, and demographic information. Examples include the fact that CSF and PET scans may detect amyloid- accumulation in the brain years before any outwardly noticeable structural abnormalities associated with the illness manifest. However, MRI scans have shown a higher sensitivity to changes that occur after the onset of symptoms\cite{zhang2012multi}. Medical history, including age and level of education, cognitive screening instruments (CSs) such as the Mini-Mental State Examination (MMSE), the Alzheimer's Disease Assessment Scale for Cognition (ADAS-cog), the Frontotemporal Dementia Questionnaire (FTDQ), and characteristics obtained from neuropsychological tests such RAVLTs, have all been shown to be significant as predicting components for AD development\cite{duchesne2009relating}.

However, when it comes to learning numerous tasks by merging various time series modalities together, the constraints of standard ML classifiers make it difficult to effectively address these problems\cite{tabarestani2020distributed}. The predictive abilities of deep learning (DL) methods have been shown to be rather impressive\cite{suresh2017clinical, talukder2023efficient, talukder2023empowering}. This research proposes a model to detect Alzheimer's disease from brain MRI images. The VGG-16 model, comprised of 16 convolution layers and features with a uniform architecture, is an optimal choice for extracting features from images \cite{talukder2022machine}. Following the feature extraction process, the LSTM model with the FC layer is employed for image classification. The outcomes of this investigation underscore the remarkable diagnostic prowess of the proposed hybrid model, achieving an impressive 98.8\% accuracy rate in detecting instances of Alzheimer's disease.

Our primary contributions unfold as follows::
\begin{itemize}
\item Pioneering the development of a hybrid deep learning model designed to identify early signs of Alzheimer's disease. This hybrid configuration entails the utilization of VGG16 for feature extraction, LSTM for encoding these features, and fully connected layers enhanced by a softmax function for image classification.
\item Executing a comprehensive performance assessment encompassing accuracy, precision, recall, and f1-score metrics. Our scrutiny validates the efficacy of our approach, as evidenced by results surpassing existing literature with notable accuracy improvements
\end{itemize}

\section{Literature review}
This portion illustrates a high-level review of current research that has used ML and DL techniques for the analysis of medical data. In the past, researchers have used ensemble machine learning models to detect Alzheimer's disease risk factors. In order to diagnose Alzheimer's disease, Uddin et al.\cite{uddin2023novel} standardized an ensemble of machine-learning algorithms. The ensemble comprises Decision Tree, Random Forest, GaussianNB, Voting Classifier, XGBoost, and GradientBoost algorithms. The researchers employed the publicly accessible series of imaging studies (OASIS) dataset to train the model, assessing its precision, recall, F1 score, and accuracy. When applied to the Alzheimer's Disease (AD)-specific ADANI dataset, the voting classifier showed a maximum validation accuracy of 96

Many researchers have attempted to develop an Alzheimer's disease-detecting MRI-based CNN model. L. F. Samhan et al.\cite{samhan2022classification} used the concept of a deep-learning model to predict Alzheimer's disease. They used a CNN model named VGG16 and the ADNI dataset. They got 97\% accuracy from their model. What is more, K. Yang et al.\cite{yang2021detection} have used a graph-regularized CNN model to identify Alzheimer's disease. The dataset, they have used, has 4 different classes. They got 97.8\% accuracy from their model. Choi et al.\cite{choi2020combining} proposed a model in DCNN to find optimal fusion weight. As an input, they combine multiple MRI projects. M. Kavitha et al.\cite{kavitha2019multi} put out a novel approach that integrates U-net-like convolutional neural networks (CNNs) for the purpose of detecting and categorizing Alzheimer's Disease. The researchers used the Alzheimer's Disease Neuroimaging Initiative (ADNI) dataset, which consisted of three-dimensional magnetic resonance imaging (MRI) pictures. With CNNs, they also used multinomial logistic regression. They got an accuracy of 97\%.

\begin{table}[h]
\caption{{Related Works with Methodologies and Classes}}
\label{tab:comparision1}
\begin{center}
\begin{tabular}{p{.8cm} p{2cm} p{3cm} p{1.4cm}}
\hline
\multicolumn{1}{l}{Study}   & \multicolumn{1}{l}{Methodology}   & \multicolumn{1}{l}{No. of Class} &  \multicolumn{1}{l}{Accuracy} \vspace{1mm} \\ \hline

 \cite{yang2020review} & VGG16                     & Binary Class          & 73.46\%         \\
    ~                  & ~                         & (AD vs NC)            & ~   \vspace{1mm}  \\ 
 \cite{venugopalan2021multimodal} & Multiscale -Deep-learning & Binary Class          & 82.40\%          \\
    ~                  & ~                         & (AD vs HC)            & ~   \vspace{1mm}    \\
    \cite{pradhan2021detection} & 2D-CNN                    & Binary Class          & 89.80\%          \\
    ~                  & ~                         &  (AD vs MCI)          & ~      \vspace{1mm}  \\
    \cite{razavi2019intelligent} & Siamese Network           & Binary Class          & 92.72\%         \\
    ~                  & ~                         &  (AD vs MCI)          & ~     \vspace{1mm} \\
    \cite{saleem2022deep} & Deep Polynomial-Network   & Multi-Class           & 55.34\%         \\
    ~                  & ~                         & (AD vs NC)            & ~   \vspace{1mm}  \\
    \cite{zaabi2020alzheimer} & Deep CNN                  & Three Class           & 95.73\%         \\
    ~                  & ~                         &  (NC, EMCI, LMCI)     & ~    \vspace{1mm}     \\
    \cite{folego2020alzheimer} & Ensemble model            & Four Class            & 94.03\%         \\
    ~                  & ~                         & (AD vs HC)            & ~   \vspace{1mm}  \\
    \cite{Kaggle_dataset} & FreeSurfer                & Four Class            &  75\%           \\
    ~                  & ~                         &  (AD, NC, EMCI, LMCI) & ~  \vspace{1mm}   
    \\ \hline

\end{tabular}
\end{center}
\end{table}

Yang et al.\cite{yang2020review} introduced an automated model for detecting Alzheimer's Disease, leveraging the Inception V4 CNN architecture. Their approach utilized the OASIS dataset and involved customizing Inception B and C modules. The model underwent training and testing phases, resulting in an accuracy of 73.75\%. In a separate study, Venugopalan et al.\cite{venugopalan2021multimodal} presented a technique focused on denoising MRI scans using autoencoders to extract inherent features from the provided data. Their experimentation encompassed 220 patients, with a training dataset of 503 MRI images. Notably, this endeavor yielded an accuracy of 78\%. 
A. Pradhan et al.\cite{pradhan2021detection} developed a methodology that uses the VGG19 model. They used 6000 images with 4 different classes mild, moderate, very mild, and non-demented. They also used DenseNet For image classification. After comparing they found that VGG19 performs better than DenseNet. 


\section{Methodology}
\subsection{Adopted Base Model Architecture}
\begin{itemize}
    \item\textbf{VGG-16 Architecture design:}
     The VGG16 model comprises 16 convolution and 5 max-pool layers. The supplied picture is scaled to 224*224*3\cite{vgg16}. 
     
     \begin{figure}[h]
     \centering
     \includegraphics[width=.5\textwidth]{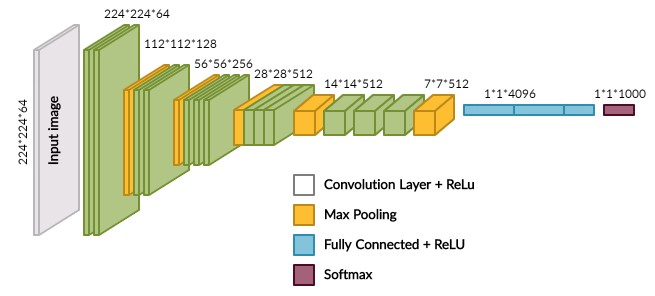}
     \caption{VGG16 Architectur}
     \label{fig:my_label}
     \end{figure}
     
     There are 5 blocks. Each block contains 2–3 convolution layers and 1 max-pooling layer. Every convolution layer has a stride(2,2) max-pool layer and a 3*3 filter with the same padding. After (7,7,512), a feature map is created. Then, it flattens the layer to a dimensional array with a feature vector of (1,1,25088). Output is delivered to the hidden layer, which is the ReLu activation layer. This layer reduces vanishing gradients and speeds learning. After that, the softmax activation layer normalizes the input by a probability distribution and classifies the pictures.

\item\textbf{LSTM Architecture design:}
LSTM stands for Long Short Term Memory. It has a feedback connection. It has the ability to store any information for a long time. LSTM mainly works based on present input by considering previous output. After that, the output is stored in the short term memory of LSTM. Cell state is the key of LSTM\cite{lstm_colah}. The Long Short-Term Memory (LSTM) mechanism is responsible for determining which information is to be discarded from the cell state.

\begin{figure}[h]
    \centering
    \includegraphics[width=.5\textwidth]{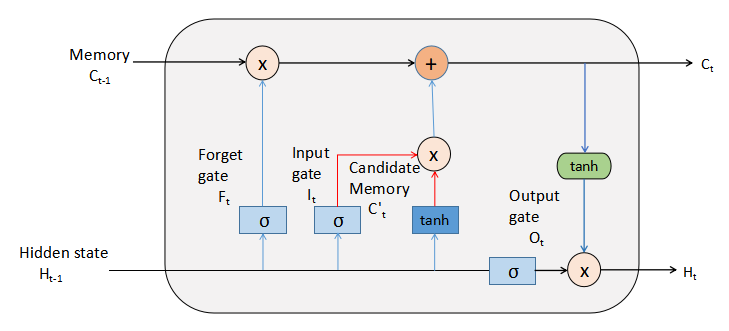}
    \caption{LSTM Architecture}
    \label{fig:my_label}
\end{figure}

LSTM works with three gates and two functions\cite{lstm_gate}. Forget gate decides which information should be forgotten from long-term memory or to keep it. Input gates get new information and specify its importance by comparing it with the previous information. The output gate gives the output based on the previous two gates. These gates use the sigmoid function and tanh function.

\end{itemize}

\subsection{Proposed Hybrid Model}
A deep learning-based hybrid model is proposed here. VGG-16 is used for feature extraction \cite{islam2023deep}. After that, LSTM is used to capture and leverage sequential context within data. The proposed hybrid model contains 13 convolution layers, 5 pooling layers, 1 LSTM layer, and the fully connected layers, represented in Fig. 3. 

\begin{figure*}[]
\centering
\subfloat[Flow Diagram]{\includegraphics[scale=.35]{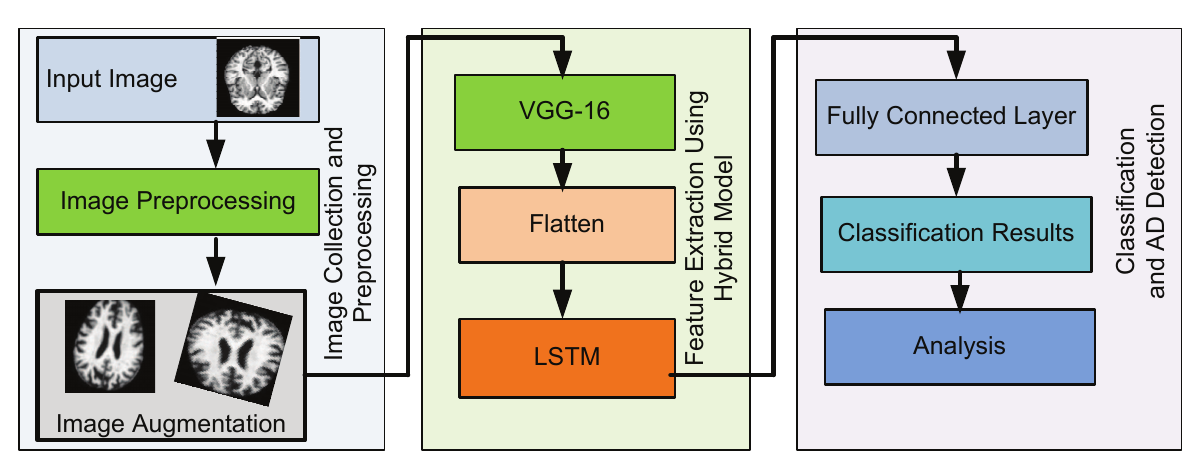}}\hspace{0.3cm}
\subfloat[Details of the Proposal]{\includegraphics[scale=.35]{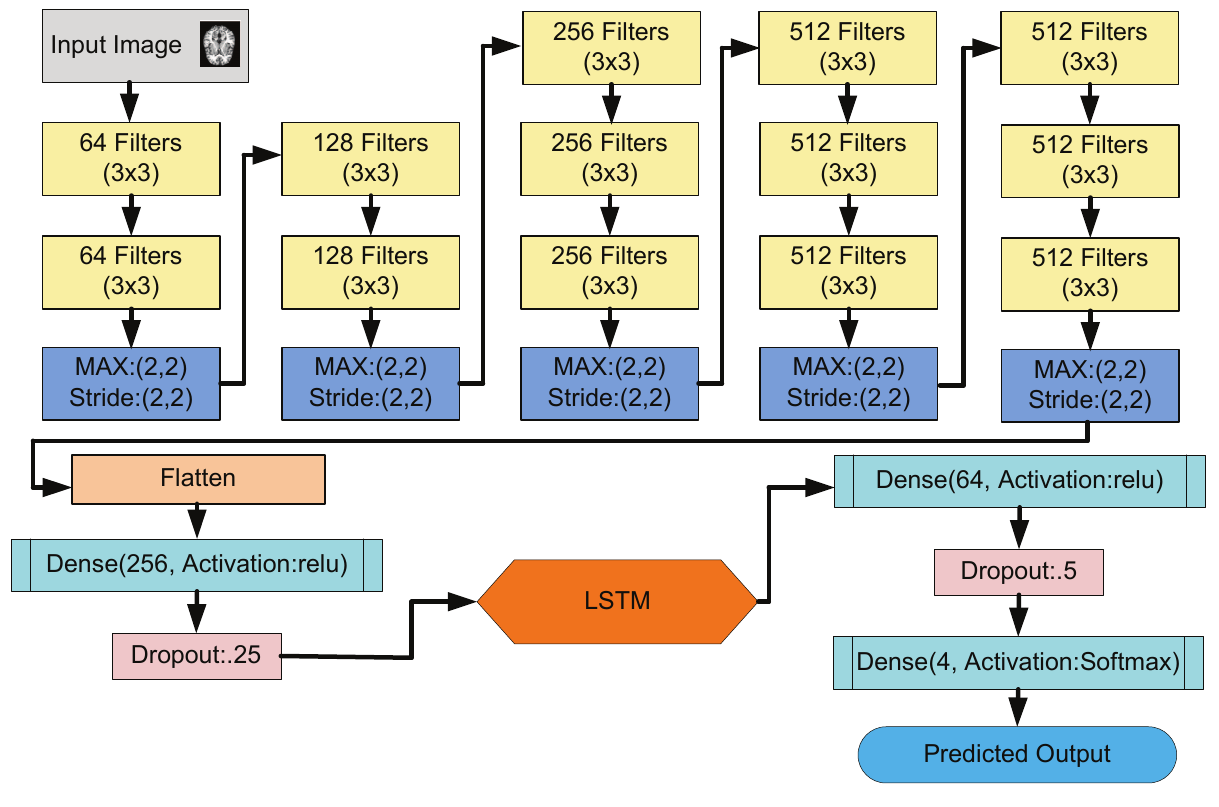}}
\caption{Proposed Hybrid Model for Alzheimer's Detection}
\label{fig:proposed}
\end{figure*}

\begin{figure}[h]
    \centering
    \includegraphics[width=.4\textwidth]{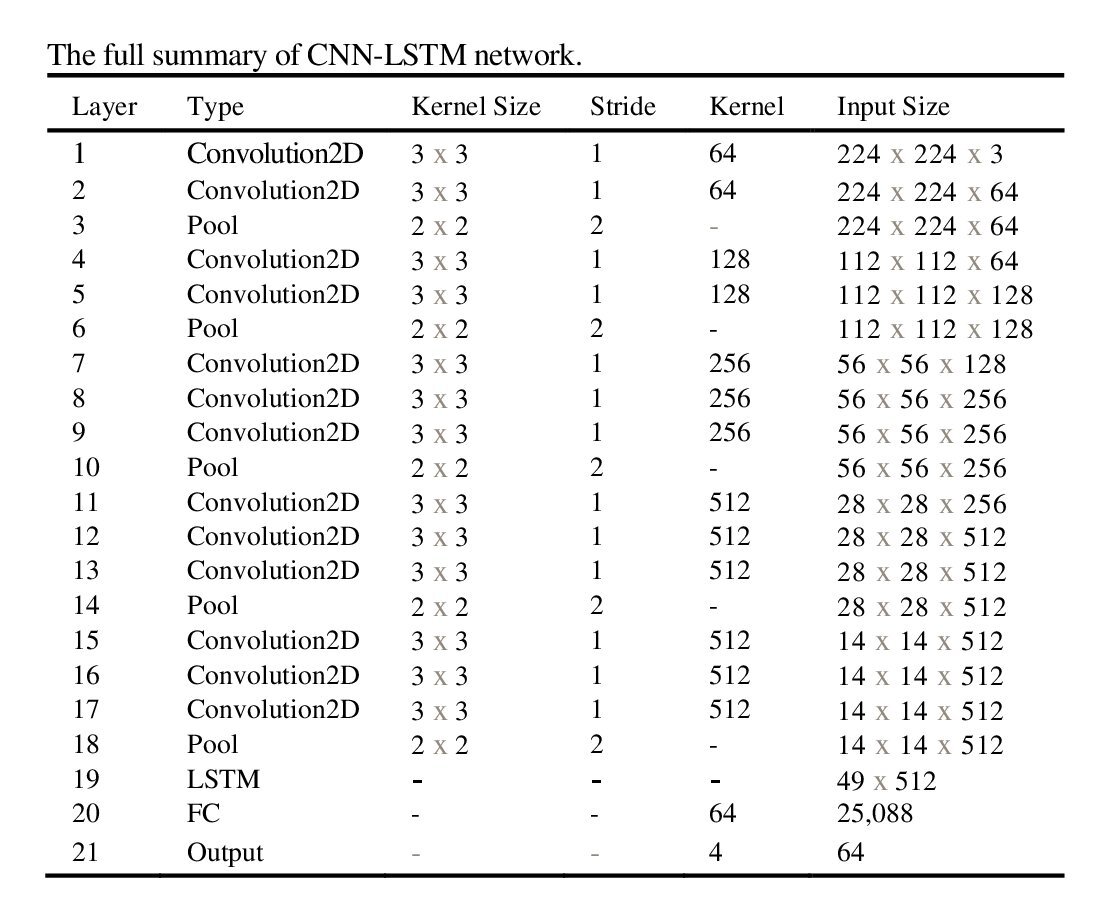}
    \caption{Summary of Proposed System}
    \label{fig:summary}
\end{figure}

Convolution layers form the characteristics of each input image. All the convolutional layers use a 3*3 kernel size. The convolutional layer also contains the ReLu activation function. Because it makes all negative values zero. The process continues until it goes through all the images. VGG16 model follows 5 blocks of convolutional layer. Max pooling is used after every block on the convolutional layer. Every max pooling layer has the same pool size and strides. The convolution and pooling layer breaks up the images into features and analyzes them. The pooled feature maps are transferred to the flattened layer. The flattened layer takes the input, makes it a single column, and transfers it to the LSTM layer. \\\\ 
The LSTM network stores and retains all the extracted characteristics obtained from the Convolutional Neural Network (CNN). The completely connected layer is supported by this process in order to make the final determination of classification. The first fully connected layer uses the LSTM's inputs as weights in order to provide a prediction about the label. The fully connected (FC) layer subsequently generates the ultimate probability for the final labels. Fig. \ref{fig:summary} presents a concise overview of the proposed concept.

\subsection{Dataset}\label{AA}
A public dataset is formed from different sources by collecting the MRI images of the brain. Data was collected from the kaggle\cite{Kaggle_dataset}. The dataset has 4 different classes.
\begin{enumerate}
    \item Mild demented
    \item Moderate demented
    \item Non demented
    \item Very mild demented 
\end{enumerate}

\begin{figure}[h]
    \centering
    \includegraphics[width=0.4\textwidth, inner]{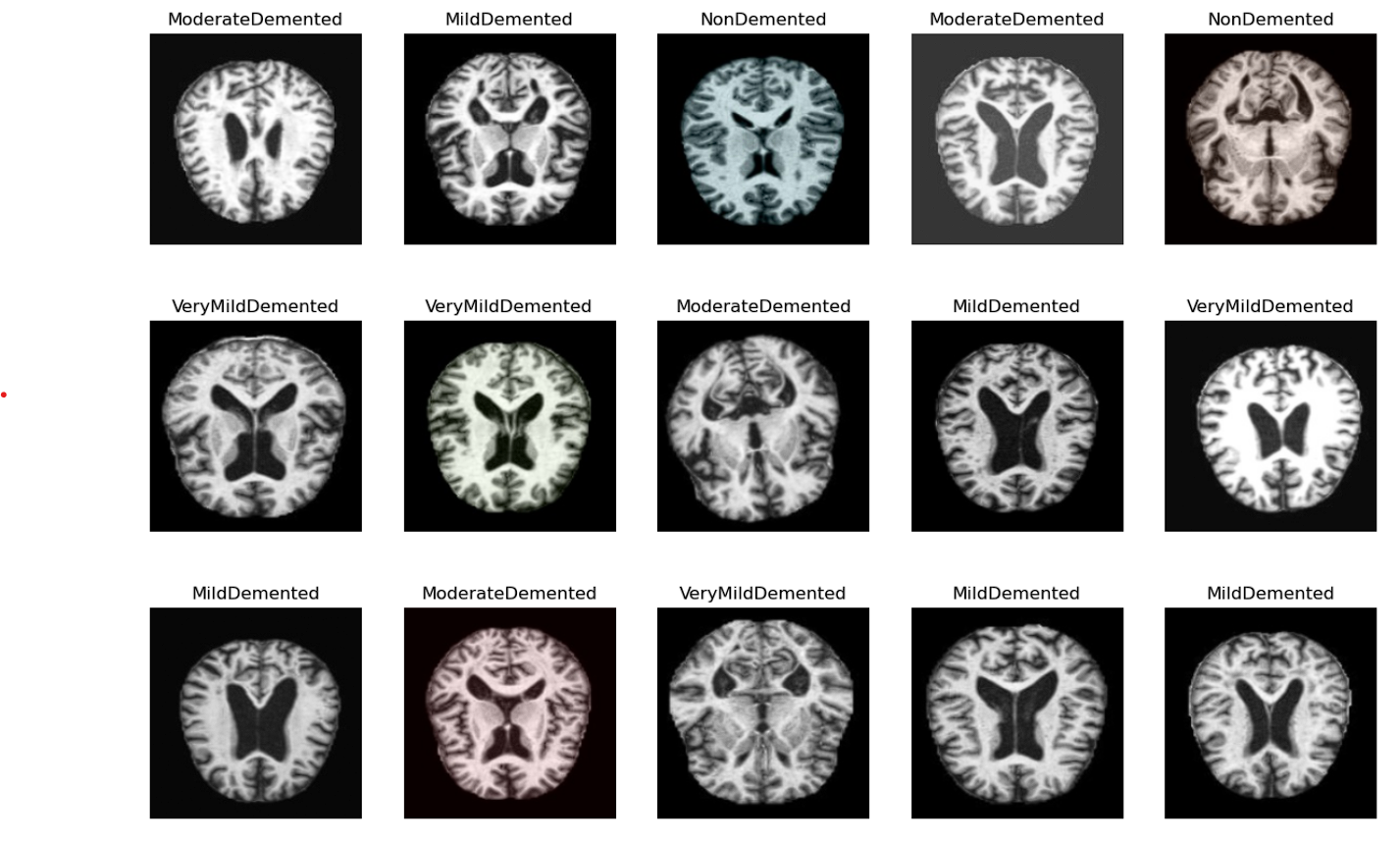}
    \caption{Data Show}
    \label{fig:my_label}
\end{figure}

Around 6400 MRI images are used to train and test the model. Of which, 896 images are mild demented, 64 images are moderate demented, 3200 images are very mild demented, and 2240 images are non demented. Fig.5. represents the output of the dataset.

\subsection{Data Preprocessing and Parameters}\label{BB}
Before applying the dataset to the machine learning algorithm, the dataset should be preprocessed. So, the images of the dataset are resized in preprocessing technique. 150 rotation and nearest fill mode is used as data augmentation technique. The training dataset comprises 80\% of the total accessible data, with the remaining 20\% reserved for testing purposes. The suggested model's input size is characterized by dimensions of 224 pixels in width, 224 pixels in height, and 3 channels representing the RGB color space. Our model was trained using 32 batches, 100 epochs, and a 0.001 learning rate. Categorical cross-entropy is a loss function.
Adam optimizer was utilized for the stochastic gradient. The hybrid deep model for Alzheimer's disease detection worked as planned.



\section{Result and discussion}

\subsection{Environment Setup}

TensorFlow Keras package classifies, and the model runs on an AMD Ryzen 7 with 32 GB RAM. All computational tasks in the proposed system are done in Python.
Our technique is evaluated using accuracy, precision, recall, F1-measure, specificity, and a confusion matrix. True Negative (TN), True Positive (TP), False Negative (FN), and False Positive (FP) represent the model's performance in the confusion matrix. The performance metrics utilized in this study are listed below:
\textbf{Accuracy : }

\begin{equation}
    Accuracy= \frac{TP + TN}{TP + TN + FP + FN}
\end{equation}

\textbf{Precision: }
\begin{equation}
     Precision = \frac{TP}{TP+FP}
\end{equation}

\textbf{Sensitivity : }
\begin{equation}
     Recall = \frac{TP}{FN + TP}
 \end{equation}

\textbf{Specificity : }
\begin{equation}
     Recall = \frac{TN}{TN + FP}
 \end{equation}
 
 \textbf{F1 measure : }
 \begin{equation}
    F_1 = 2 \times \frac{Precision \times Recall}{Precision + Recall}
\end{equation} 
Precision measures how many positive samples were actual, whereas accuracy measures how successfully the classifier predicted classes. Specificity measures a model's ability to predict negative outcomes, whereas sensitivity measures its ability to identify positive data. When false negatives and positives are included, the F1-score is a harmonic mean of sensitivity and accuracy \cite{talukder2023dependable, talukder2024machine, talukder2024mlstl, talukder2024securing}.


\subsection{Result analysis}



\begin{figure}[!ht]
    \centering
    {\includegraphics[scale=.4]{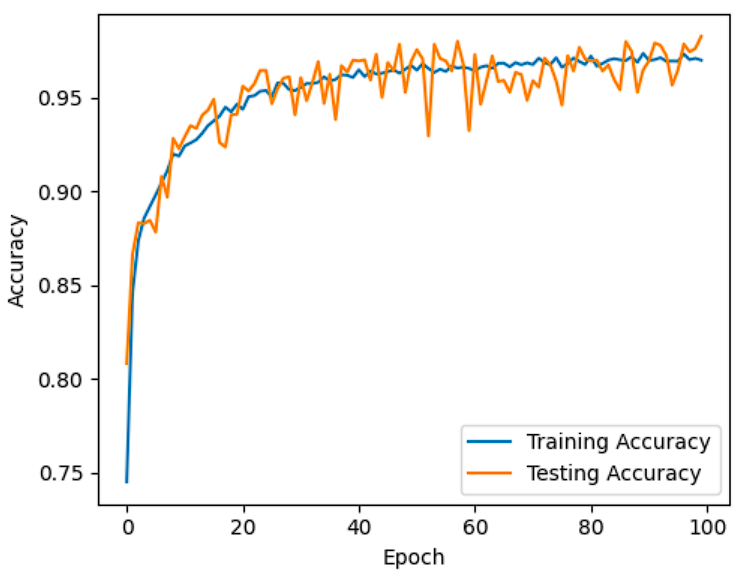}}
    \caption{Training and Testing Accuracy Curve}
    \label{fig:train}
\end{figure}

\begin{figure}[!ht]
    \centering
     {\includegraphics[scale=.4]{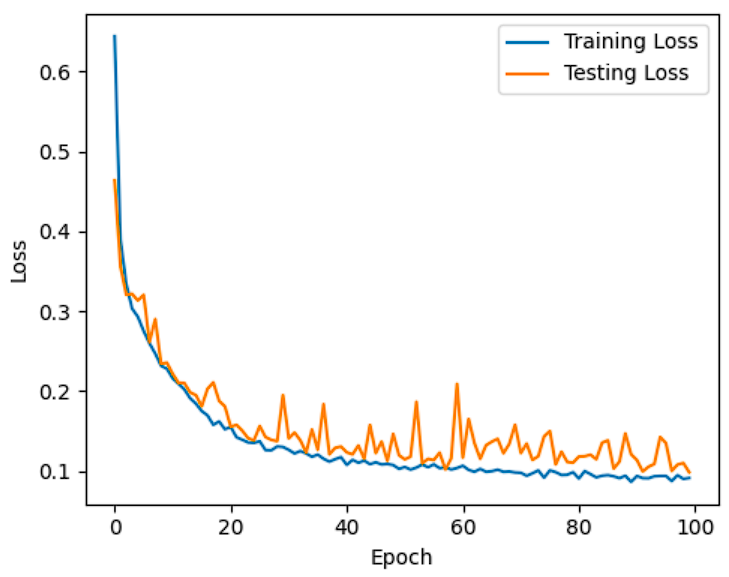}}
    \caption{Training and Testing Loss Curve}
    \label{fig:loss}
\end{figure}

Our contribution involves the development of a hybrid model, combining CNN and LSTM, to effectively detect and classify Alzheimer's disease. Detailed accuracy and loss trends throughout both training and testing stages are depicted in Figures \ref{fig:train} and \ref{fig:loss}, respectively.


\begin{figure}[!ht]
    \centering
    {\includegraphics[scale=.4]{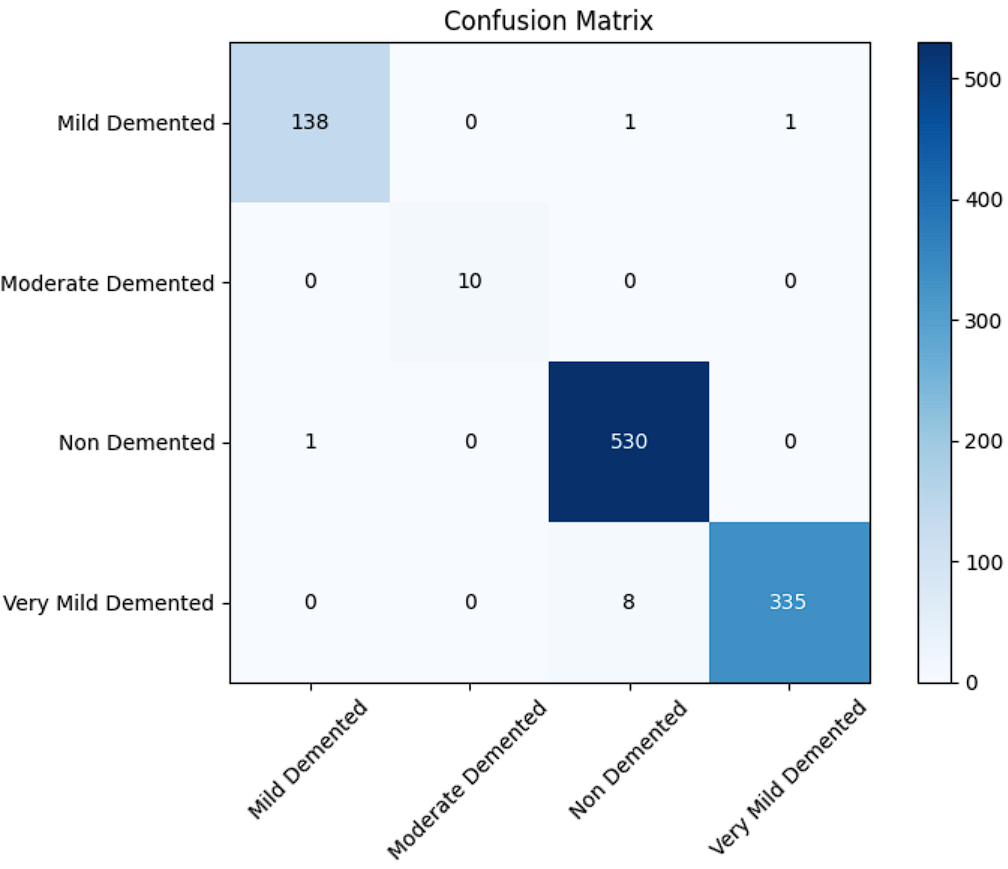}}
    \caption{Confusion Matrix}
    \label{fig:experiment1}
\end{figure}

The confusion matrix of our hybrid model is shown in Fig. \ref{fig:experiment1}. The rows define the actual classes, while the columns represent the predicted class. The number of cases that were assigned to each category is shown in the matrix cells.
138 instances were classified as Mild demented correctly, while only 2 instances were not classified accurately. In the case of Moderate demented, 10 instances were classified correctly, while no instances were falsely classified. Only 1 instance of Non-Demented was falsely classified, while 530 instances were correctly classified. For Very mild demented, 335 instances were classified correctly, while only 8 were classified as Non demented.


\begin{figure}[h!]
      \centering
          \includegraphics[width=.35\textwidth]{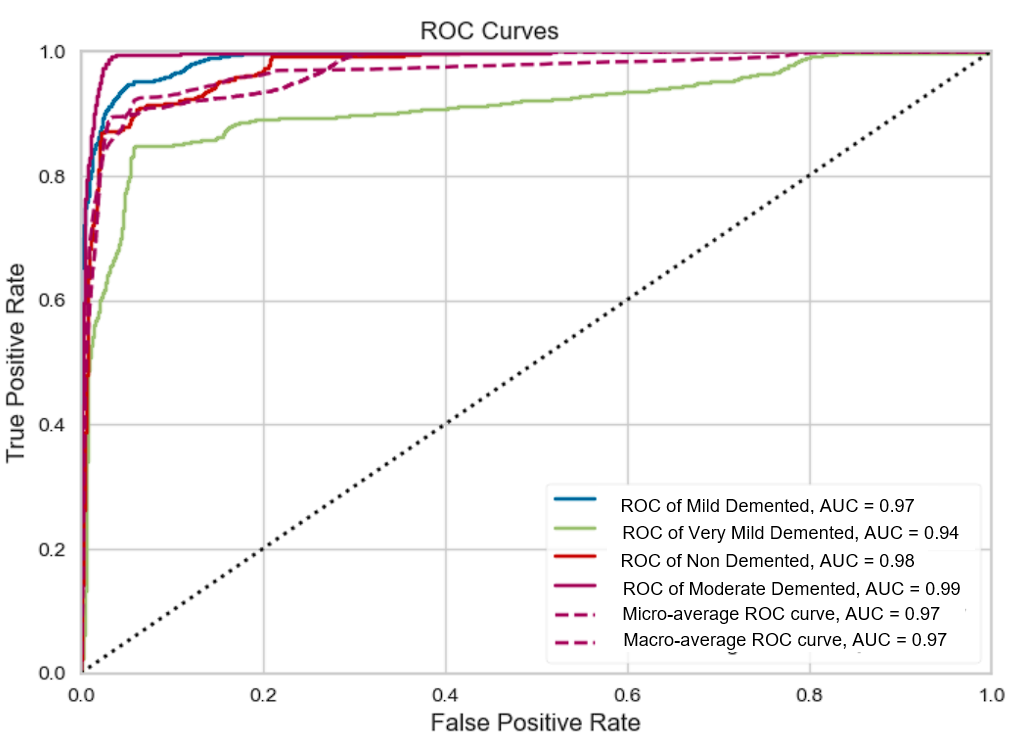}
      \caption{ROC-AUC curve}
      \label{fig:roc}
\end{figure}

Fig. \ref{fig:roc} illustrates the ROC curve, which was constructed by evaluating the true positive ratio against the false positive ratio across various precision thresholds. The ROC curve vividly showcases the remarkable performance of the hybrid model proposed in this study, yielding area under the curve values of 0.97 for the Demented class, 0.94 for the Mildly Demented class, 0.98 for the Non-Demented class, and an impressive 0.99 for the Moderately Demented class.

\begin{figure}[h]
      \centering
          \includegraphics[scale=0.25]
          {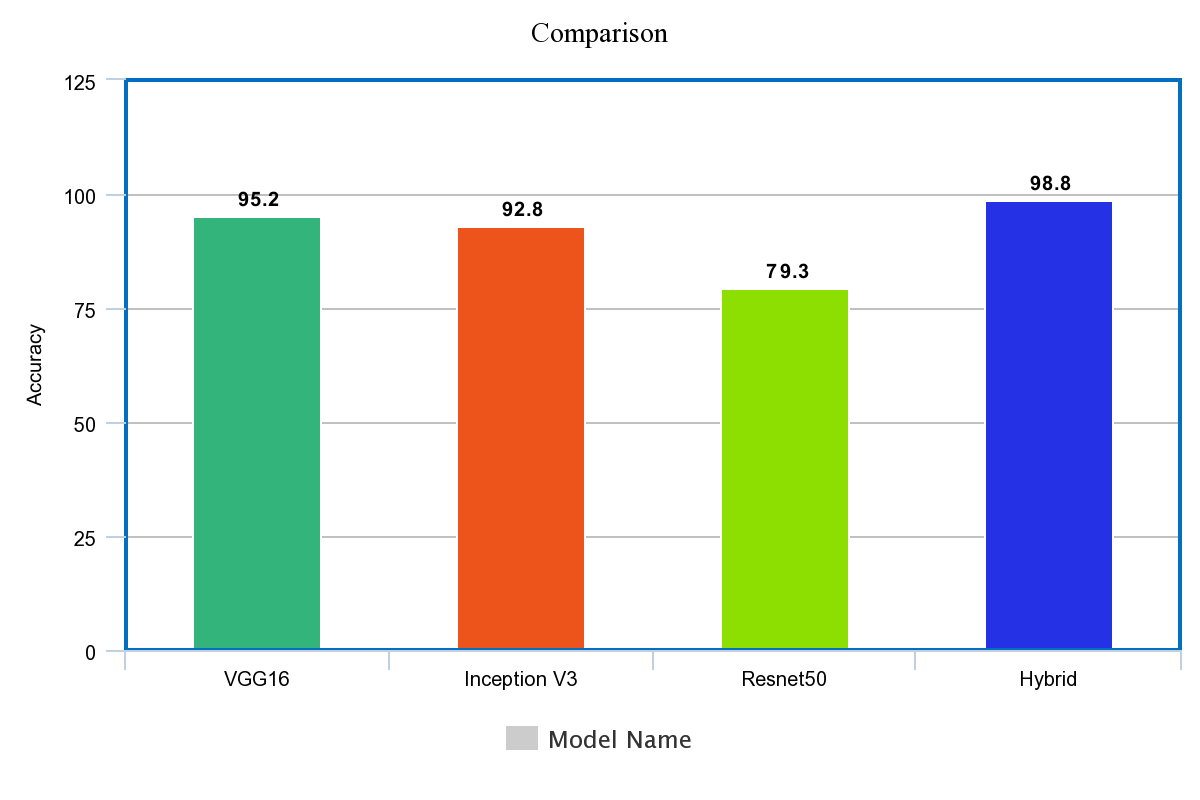}
      \caption{Comparison between Traditional CNN Model}
      \label{fig:my_label}
\end{figure}
 The performance of our hybrid model surpasses that of several cutting-edge approaches, as illustrated in Fig. \ref{fig:my_label}. Our model achieves an accuracy of 98.8\%, with sensitivity at 100\%, a precision rate of 98.08\%, a specificity of 76\%, and a f1-score of 99.03\%.

\subsection{Discussion}
Table \ref{tab:comparision1} represents the comparisons of different models proposed for classifying Alzheimer's disease using MRI images by many authors.  L. F. Samhan et al. \cite{samhan2022classification} used VGG16 model on ADNI dataset, achieving 97\% accuracy while
M. Kavitha et al.\cite{kavitha2019multi} have the same accuracy as the CNN-Logistic model. Meanwhile, T. J. Saleem et al.\cite{saleem2022deep}, and G. Folego et al.\cite{folego2020alzheimer} have lesser accuracy on the same dataset with a different approach which is 67\% and 52.3\% respectively. K. Yang et al.\cite{yang2021detection} used the VGG19 model on 3210 images with 4 different classes, achieving an accuracy of 97.8\%.

\begin{table}[h]
\caption{{Comparison among different related working models and proposed model}}
\label{tab:comparision1}
\begin{center}
\begin{tabular}{p{3cm}  p{1.5cm}  p{1.2cm} p{.9cm}}
\hline
\multicolumn{1}{l}{Authors}   & \multicolumn{1}{l}{Number of class}   & \multicolumn{1}{l}{Methodology} &  \multicolumn{1}{l}{Accuracy} \\ \hline

 L. F. Samhan et al. \cite{samhan2022classification} & ADNI dataset &	VGG16  & 97\%  \\ 

 K. Yang et al. \cite{yang2021detection}	& 3210 images + 4 classes	& VGG19	& 97.8\%    \\
 
 Z. Cui et al. \cite{cui2019alzheimer} & 2400 images + 2 classes & Inception V3 & 85.7\%     \\

 
 V. Patil et al. \cite{article} &	223 images & DenseNet &	96.4\%    \\ 

 E. A. Mohammed et al. \cite{yang2020review}	& OASIS dataset	& Inception V4	& 73.75\%   \\ 

 J. Venugopalan et al. \cite{venugopalan2021multimodal}  &	503 images &	CNN &	78\%  \\

 A. Pradhan et al. \cite{pradhan2021detection}	& 6000 images + 4 classes &	VGG16	& 94\%  \\

 F. Razavi et al. \cite{razavi2019intelligent} 	& 51 images &	S Filter + Regression	& 98.3\%  \\
 
 T. J. Saleem et al. \cite{saleem2022deep} 	& ADNI dataset & DNN	& 67\%   \\

 M. Zaabi et al. \cite{zaabi2020alzheimer}		& 4870 images + 2 classes & CNN + Transfer Learning	& 92.81\% \vspace{1.1mm}\\

 G. Folego et al. \cite{folego2020alzheimer}	& ADNI dataset &		LeNet-5 & 52.3\% \\

 Feng C et al. \cite{feng2019deep}	& ADNI dataset &		3D-CNN
and FSBi-LSTM & 94.82\% \\

 Dua M et al. \cite{dua2020cnn}	& OASIS dataset &		CNN + RNN + LSTM & 92.22\% \\

 \textbf{Proposed model}		& \textbf{6400 images + 4 classes} &	\textbf{VGG + LSTM}	& \textbf{98.8\%} \\
 \hline

\end{tabular}
\end{center}
\end{table}

A. Pradhan et al.\cite{pradhan2021detection} applied VGG16 on 6000 images with 4 different classes and achieved an accuracy of 94 \%. On the OASIS dataset, E. A.
Mohammed et al.\cite{yang2020review} applied the inception V4 model and achieved an accuracy of 73.75\%. M. Zaabi et al.\cite{zaabi2020alzheimer}, applied a combination of CNN and Transfer learning on 4870 images with 2 classes, which achieved an accuracy of 92.81\% accuracy. No other author has applied their model on a larger dataset than 3000 images with more classes. Some extraordinary combinations have been noticed in this field of study. In the study of Feng C et al. \cite{feng2019deep}, a decent 94.82\% accuracy was gained using a combination of 3D-CNN and FSBi-LSTM models. On the other hand, Dua M et al. \cite{dua2020cnn} applied a combination of DL models such as CNN, RNN, and, LSTM on the OASIS dataset for achieving 92.22\% accuracy.

In comparison, the model proposed in this study, which is a hybrid approach, achieved the highest accuracy of 98.8\% on 6400 images with 4 classes, which is a larger dataset than all of them and also outperforms all other models' accuracy. The potential of this model to diagnose Alzheimer's disease is promising.

\section{Conclusions and Future Work}
The suggested deep learning model improves upon previous methods for detecting and classifying Alzheimer's disease in MRI scans. It can identify photos with ease by automatically extracting characteristics from them. 
The suggested model beat many previous models and achieved impressive results, including a 98 \% accuracy rate, a precision rate of 98.8 \%, a specificity rate of 76 \%, a f1-score of 99.3 \%, and a sensitivity rate of 100 \%.
The model will be improved in the future by fine-tuning it to achieve higher precision. The suggested model will also be tested on a variety of datasets.

\section{Acknowledgement}
This research was jointly supported by the ICT Innovation Fund, ICT division, Bangladesh; Title "Hybridized Convolutional Neural Networks and Long Short-Term Memory for Improved Alzheimer’s Disease Diagnosis from MRI Scans", Year 2022-2023 and Jagannath University Research Grant, Year 2022-2023.




\bibliographystyle{IEEEtran}
\bibliography{bibfile}

\end{document}